\documentclass[a4paper]{spie}  

\usepackage[displaymath, mathlines]{lineno}
\usepackage{xcolor}
\newcommand{\revA}[1]{{\color{black}#1}}
\newcommand{\revB}[1]{{\color{black}#1}}

\usepackage{amsmath,amsfonts,amssymb}
\usepackage{graphicx}
\usepackage[colorlinks=true, allcolors=black]{hyperref}

\title{Efficient Bayesian inversion for shape reconstruction of lithography masks}

\author[a,b]{Nando Farchmin}
\author[c,d]{Martin Hammerschmidt}
\author[c,d]{Philipp-Immanuel Schneider}
\author[a]{Matthias Wurm}
\author[a]{Bernd Bodermann}
\author[a]{Markus B\"ar}
\author[a]{Sebastian Heidenreich}
\affil[a]{Physikalisch-Technische Bundesanstalt, Braunschweig and Berlin}
\affil[b]{Technische Universit\"at Berlin, Institute of Mathematics}
\affil[c]{JCMwave GmbH}
\affil[d]{Zuse Institute Berlin}

\authorinfo{E-mail: nando.farchmin@ptb.de}

\DeclareMathOperator{\relstd}{\mathrm{rel-}2\sigma}
 
\begin{document} 
\maketitle

\begin{abstract}
  {\bf Background:}
  Scatterometry is a fast, indirect and non-destructive optical method for
  quality control in the production of lithography masks. 
  To solve the inverse problem in compliance with the upcoming need for
  improved accuracy, a computationally expensive forward model has to be
  defined which maps geometry parameters to diffracted light intensities.
  \newline
  {\bf Aim:}
  To quantify the uncertainties in the reconstruction of the geometry
  parameters, a fast to evaluate surrogate for the forward model has to be
  introduced.
  \newline
  {\bf  Approach:}
  We use a non-intrusive polynomial chaos based approximation of the forward
  model which increases speed and thus enables the exploration of the posterior
  through direct Bayesian inference.  Additionally, this surrogate allows for a
  global sensitivity analysis at no additional computational overhead.
  \newline
  {\bf Results:}
  This approach yields information about the complete distribution of the
  geometry parameters of a silicon line grating, which in return allows to
  quantify the reconstruction uncertainties in the form of means, variances and
  higher order moments of the parameters.
  \newline
  {\bf Conclusion:}
  The use of a polynomial chaos surrogate allows to quantify both parameter
  influences and reconstruction uncertainties. This approach is easy to use since
  no adaptation of the expensive forward model is required.
\end{abstract}


\section*{ACKNOWLEDGEMENTS}

This project has received funding form the German Central Innovation Program
(ZIM) No. ZF4014017RR7.

\section{INTRODUCTION}
\label{sec:intro}

Scatterometry is an optical scattering technique frequently used for the
characterization of periodic nanostructures on surfaces in semiconductor
industry  (determination of critical dimensions)
\cite{HT04,mack2008fundamental,scholze2013comparison,henn2014improved}. In
contrast to other techniques like electron microscopy, optical microscopy or
atomic force microscopy, scatterometry is a non-destructive and indirect
method. In the last decades both feature sizes and the required measurement
uncertainty decreased continuously, hence advanced scatterometry techniques are
required. Recently, deep ultraviolet (DUV) scatterometry \cite{WBBR11,ABB+15,
wurm2017metrology}, extreme ultraviolet (EUV) scatterometry
\cite{raymond1995metrology, HT04}, imaging scatterometry
\cite{madsen2016imaging} and combinations with white light interferometry
\cite{paz2012solving} are developed. In these approaches, geometrical
parameters and associated uncertainties can be determined from diffraction
patterns by solving a statistical inverse problem  \cite{germer2009developing}.
For an overview on the metrology of surfaces in semiconductor industry see
\cite{orji2018metrology} and references therein. 

We emphasize that scatterometry is an integral measurement method, which means
that information of variances within the probe are lost due to an averaging
over the spot size of the beam. These parameter variations typically lead to a
broadening of the diffracted beam. This stochastic effect was not taken into
account by the model of the line structure used in this work. Instead the
diffraction efficiencies were calculated from the integral over the whole beam.



The inverse problem of scatterometry is in general ill-possed and
regularization techniques have to be applied. The geometry is typically
parametrized and sought-after parameters are obtained by weighted least squares
minimization \cite{el2011performance}, with weights derived
directly from uncertainties in the measurements.  However, the quality of these
weights depends highly on \revB{the measurements used} and itself influences the
reconstruction results of the geometry parameters\cite{HGSWEB12}.  \revB{An alternative
approach} is to apply a maximum likelihood estimate, which introduces a
likelihood function based on an error model and optimizes \revB{weighting terms} as hyper
parameters instead of using predefined values.  Based on the same principle
but additionally including some prior knowledge is the maximum posterior
approach, which is a state of the art method in parameter
reconstructions\cite{HWS+17}.  In the above frameworks, uncertainties are
typically obtained from the Fischer information or covariance matrix , which relies on an
assumed shape of the posterior.  However, the shape of the posterior is
generally unknown, hence this can lead to significant errors in the estimation of
uncertainties if the actual posterior shape differs from the assumed one.

The Bayesian approach allows to integrate prior knowledge \cite{HGWBB15} and
approximates the probability density function of the geometry parameters
independent of any shape assumptions. Uncertainties obtained by employing
Bayesian inference are thus much more \revB{robust}.
On the other hand it requires a large number of evaluations of the forward
model which is not feasible for expensive computations as in the case of
scatteromery.  To obtain a surrogate model that mitigates the computation
time, we employ a polynomial chaos expansion, that is an expansion into an
orthonormal polynomial basis in the parameter space to approximate the forward
model with a global polynomial \cite{Sud08, Xiu09}.  We additionally show how
this surrogate allows for a Bayesian approach to the inverse problem.

\revA{
In a recent publication, it has been demonstrated that a surrogate of a forward
model for scatterometry based on a polynomial chaos expansion enables Bayesian
inversion and the use of Markov Chain Monte Carlo (MCMC) sampling. In this
approach, cubature rules on sparce grids of Smolyak type are used to determine
the expansion coefficients and to construct the surrogate model
\cite{HGHEB14,HGB18}. However, cubature rules and sparce grids are adapted to
the stochastic distributions chosen and less accurate for correlated stochastic
input parameters. In the present work we used optimal linear regression to
obtain the coefficients of the polynomial chaos expansion. This novel approach
uses optimal sampling points, is much more flexible and well suited for
extensions to adaptive systems.
}

\begin{figure} [ht]
  \begin{center}
    \begin{tabular}{c} 
      \includegraphics[width=.8\linewidth]{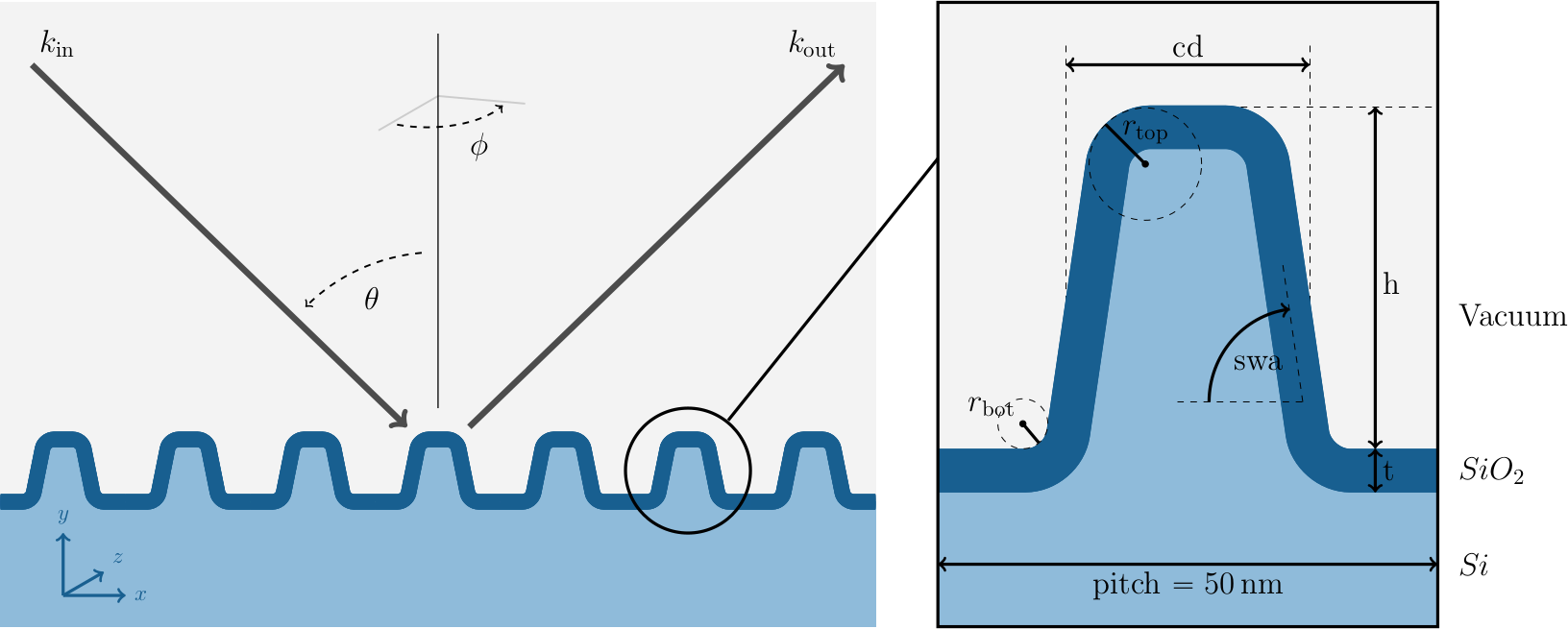}
    \end{tabular}
  \end{center}
  \caption{\label{fig:scat_setup} 
    Cross section of the photomask with description of the stochastic
    parameters. The dimensional parameter vector is given by
    $\xi=(h,\mathrm{cd},\mathrm{swa},t,r_\mathrm{top},r_\mathrm{bot})$.  The
    pitch of the computational domain, i.e. the period is fixed to
    $50\,\mathrm{nm}$.
    }
\end{figure} 

In this paper, we determine the geometry parameters of a photomask that consists
of multilayered, periodic, straight absorber lines of two optically different
materials. The period of the line structure (pitch) is $50\,\mathrm{nm}$ and
the geometry parameters of interest are the height of the line $h$, the width
at the middle of the line (critical dimension) $\mathrm{CD}$, the sidewall
angle $\mathrm{SWA}$, the silicon oxide layer thickness $t$ and the radii of
the rounding at the top and bottom corners of the line $r_\mathrm{top}$ and
$r_\mathrm{bot}$, respectively. A cross section of the geometry for one period
of the structure is depicted in Fig.~\ref{fig:scat_setup}. The photomask was
illuminated by a light beam of wavelength $\lambda=266\,\mathrm{nm}$ (DUV) for
different angles of incidence $\theta=3^\circ,5^\circ,\dots,87^\circ$ for
perpendicular ($\phi=0^\circ$) and parallel ($\phi=90^\circ$) orientation of
the beam with respect to the grating structure as well as S and P polarization.
The refractive indices used are $n_\mathrm{si}=1.967+4.443i$ for silicon,
$n_\mathrm{ox}=1.7212+0.113i$ for the top oxide layer and $n_\mathrm{air}=1.0$
for air. All the materials are assumed to be isotropic and non-magnetic
($\mu_r=1.0$). 


In the next sections, we will proceed as follows. First, we introduce the
forward model of the problem. Second, we describe a non-intrusive method to
build a polynomial chaos based surrogate by utilizing an optimal sampling
strategy.In Section~\ref{sec:bayes} and \ref{sec:results} we apply Bayesian
inversion to our surrogate. Finally, we estimate the posterior distribution
from measurement data and compare the results.

\section{FORWARD MODEL}
\label{sec:forward_model}

 In principle, the propagation of electromagnetic waves is described by
Maxwell's equations, but for our simple grating geometry and in the time-harmonic
case, Maxwell's equations reduce to a single second order
partial differential equation \cite{Mon03, HWS+17},

\begin{align}
  \label{eqn:Maxwell}
  \operatorname{\nabla\times} \mu(r)^{-1} \operatorname{\nabla\times} E(r)
  - \omega^2 \varepsilon(r) \, E(r) = 0.
\end{align}

Here, $\varepsilon$ and $\mu$ are the permittivity and permeability, $r$
is the spatial coordinate  and $\omega$ is the frequency of the incoming beam.
We employ the finite element method (FEM) \cite{Mon03} implemented in the
JCMsuite software package to discretize and solve the corresponding scattering
problem formulation on a bounded computational unit cell in weak formulation as
described in \cite{Pomplun2007}. This formulation yields a splitting of the
complete $\mathbb{R}^n$ into an interior domain hosting the total field
(incident and scattered) and an exterior domain where only the purely outward
radiating scattered field is present. Appropriate  boundary conditions are
applied at the boundary of the computational domain as depicted in
Fig.~\ref{fig:scat_setup}. As the geometry is periodic in lateral direction
Bloch-periodic boundary conditions are applied. In vertical direction the
geometry is assumed to be unbounded and thus requires the satisfaction of a
transparent boundary condition at the interface. We use an adaptive perfectly
matched layer (PML) method\cite{Berenger1994, Zschiedrich2009} to realize the
transparent boundary condition and to satisfy the radiation condition for the
scattered field. The employed vectorial FE method uses high-order polynomial
ansatz functions defined on the spatial discretization of the computational
domain. The triangulation allows to geometrically resolve the material
interfaces and the tangential continuity of the electromagnetic fields across
these interfaces is automatically enforced by the method.

The forward model is given by a map of geometry parameters onto S and P
polarization of zeroth order intensities of the scattered light. The parameters
$\xi$ used for modelling the grating geometry are depicted in
Fig.~\ref{fig:scat_setup}. The forward model is represented by the function
$f^*\colon \Omega\to\mathbb{R}^d$ such that the parameters
$\xi\in\Omega\subset\mathbb{R}^M$ are mapped to diffracted efficiencies for a
set of azimuthal angles, incidence angles and polarizations.  Each of the
$d$ components of $f(\xi)$ represents a different combination of azimuth,
incidence angle and polarization. All other experimental conditions such as
e.g. the wavelength are fixed for this forward model. In our approach, the
experimental data $y\in\mathbb{R}^d$ are modelled with the error $y_j =
f_j(\xi) + \varepsilon_j$, $j=1,\dots,d$ where
$\varepsilon_j\sim\mathcal{N}(0,\sigma_j)$ describes a normal distributed noise
with zero mean, standard deviation and error parameter $b$,

\begin{align}
  \label{eqn:noise_variance}
  \sigma_j(b) = b\, y_j,
  \qquad \text{for } b > 0.
\end{align}

Choosing $\sigma_j$ to depend on a stochastic parameter itself instead of
setting it to specific values allows for an estimation of the measurement error
based on the measurement data in the parameter reconstruction and thus
incorporates less prior knowledge.  The inverse problem is defined as the
determination of geometry parameter values $\xi$ and the error parameter (hyper
parameter) $b$ from measured efficiencies $y$.

To obtain a fast evaluation of the surrogate, the
function $f^*$ is expanded into an orthonormal polynomial basis
$\{\Phi_\alpha\}_{\alpha\in\Lambda}\subset
L^2(\Omega;\varrho)$\cite{Wie38,CM47,GS90}

\begin{align}
  \label{eqn:pce}
  f^*(\xi) \approx f(\xi) = \sum_{\alpha\in\Lambda} f_\alpha \Phi_\alpha(\xi)
  \qquad\text{with}\qquad
  f_\alpha = \int_{\Omega}^{}f(\xi)\, \Phi_\alpha(\xi) \,\mathrm{d} \varrho(\xi).
\end{align}

The finite set $\Lambda\subset\mathbb{N}_0^M$ of cardinality $P\in\mathbb{N}$
is a set of multiindices and $\varrho$ denotes the multivariate parameter
density for the parameters $\xi$.  With this surrogate the evaluation of the
model in different parameter realizations is equivalent to the evaluation of
polynomials. 

We want to emphasize, that this approach allows for a global sensitivity
analysis of the parameters at almost no additional cost 
\cite{Sob93,Sob01,,HS96,SA10,SAA+10,Sud08,FHS+19}. 

\section{OPTIMAL LINEAR REGRESSION}
\label{sec:sampling}

A simple and non-intrusive approach to compute the expansion coefficients in
(\ref{eqn:pce}) is linear regression. With the reasonable assumption that, for
an arbitrary enumeration of $\Lambda$, the residuum $\mathcal{R}(\xi) =
f^*(\xi)-\sum_{\ell=1}^{P}f_\ell \Phi_\ell(\xi)$ is a zero mean random
variable, we want to find coefficients that minimize the variance of the
residuum $\mathcal{R}$. In other words, we obtain the least-squares
minimization problem:

\begin{align}
  \label{eqn:regression_min}
  \mbox{Find coefficients }f_\ell,\ \ell=1,\dots,P\mbox{ such that}\quad
  \int_{\Omega}^{} \mathcal{R}(\xi)^2 \,\mathrm{d}\varrho(\xi) \,=\, \min.
\end{align}

To avoid the high dimensional numerical integration in
(\ref{eqn:regression_min}), we approximate the integral in a Monte Carlo sense
by

\begin{align}
  \label{eqn:integral_approx}
  \int_{\Omega}^{} \mathcal{R}(\xi)^2 \,\mathrm{d}\varrho(\xi)
  \,\approx\, \frac{1}{N} \sum_{i=1}^{N} \mathcal{R}(\xi^{(i)})^2,
\end{align}

where $\xi^{(1)},\dots,\xi^{(N)}$ are $N$ realizations of possible
geometry parameter values (see the domain column in Tab.~\ref{tab:parameter}).
Since (\ref{eqn:regression_min}) is a quadratic minimization problem, the
critical point of the first variation yields the wanted minimum.  This critical
point can be obtained by solving the linear system

\begin{align}
  \label{eqn:linear_system}
  F = (\Psi^T \Psi)^{-1}\Psi^T F^*,
\end{align}

where the matrices $\Psi$ and $F^*$ are given by $\Psi_{i\ell} =
\Phi_\ell(\xi^{(i)})$ and $F^*_i = f^*(\xi^{(i)})$. To guarantee that the
empirical Gramian $\Psi^T \Psi$ is not ill-conditioned, the number of
realizations $N$ has to be sufficiently large. The choice of sampling points is
in principle arbitrary, but Cohen and Migliorati \cite{CM17} showed, that
sampling from a specific weighted least-squares distribution leads to an
optimal (minimal) number of samples for a guaranteed well-conditioned Gramian
matrix. Hence, we set 

\begin{align}
  \label{eqn:CM_density}
  \mathrm{d}\mu = w^{-1}\mathrm{d}\varrho
  \qquad\mbox{for}\qquad
  w^{-1}(\xi) = \frac{1}{P} \sum_{\ell=1}^{P}\vert \Phi_\ell(\xi)\vert^2
\end{align}

and draw samples $\xi^{(i)}\sim\mu$. Note that $\mu$ is still a probability
measure since the polynomials $\{\Phi_\ell\}$ are orthogonal and normalized.
With this, the number of samples required to guarantee a low condition of the
Gramian reads $N / \log(N) \geq c\, P$ for some $c>0$. Here, we choose $c=4$,
motivated by the empirical results in \cite{CM17}.

Applying this optimal sampling strategy allows us to compute a surrogate
for the forward model using a minimal number of function evaluations. 
This surrogate will be employed to reconstruct geometry parameters and 
quantify the reconstruction uncertainties using Bayesian inference.

\section{BAYESIAN APPROACH}
\label{sec:bayes}

The Bayesian approach provides a statistical method to solve the inverse
problem. Following Bayes' theorem, the posterior density is given by

\begin{align}
  \label{eqn:posterior}
  \pi(\hat\xi;y) 
  = \frac{\mathcal{L}(\hat\xi;y) \pi_0(\hat\xi)}
  {\int\mathcal{L}(\hat\xi;y)\pi_0(\hat\xi)\,\mathrm{d}\hat\xi},
\end{align}

where the prior density $\pi_0$ describes prior knowledge and the likelihood
function $\mathcal{L}$ contains the information obtained from the measurement
under the assumption of a specific measurement error model.  Since
the prior density allows for expert knowledge to influence the model, it has to
be chosen appropriately not to introduce a bias on the system. We choose a
uniform prior to induce as less information as possible on the compact domains
of the geometry parameters. The computational and implementation efforts of
the Bayesian approach are higher than for the maximum likelihood or least
squares methods.  However, the posterior yields information about the
complete probability density function of the geometry parameters and is thus
more reliable for the determination of uncertainties than merely using
quantities such as mean and covariance. In addition, the combination of
the results from different measurement modalities within the Bayesian framework
assures a consistent propagation of uncertainties through all measurement
contributions\cite{silver2011bayesian,silver2014optimizing} (hybrid metrology)
in a way that the posterior for one measurement can be used as the prior for
the next.

\begin{table}[ht]
  \caption{ \label{tab:parameter}
  Estimations of parameters and uncertainties obtained from the mean value
  (mean), the double standard deviation ($2\sigma$), relative double standard
  deviation (\ref{eqn:relstd}) ($\relstd$), skewness (skew) and
  kurtosis of the posterior distribution. The domain indicates the support of
  the prior distribution chosen.
  }
  \begin{center}       
    \begin{tabular}{|p{2.1cm}p{2.3cm}p{1.8cm}p{1.8cm}p{1.8cm}p{1.8cm}p{1.8cm}|}
      \hline
      \rule[-1ex]{0pt}{3.5ex}
      parameter & domain & mean & $2\sigma$ & $\relstd$ & skew & kurtosis \\
      \hline
      \rule[-1ex]{0pt}{3.5ex}  
      $h\ /\ \mathrm{nm}$ 
      & $[43.0 , 53.0]$ 
      & $48.35$ 
      & $3.11$ 
      & $0.6224$ 
      & $-0.0518$ 
      & $2.23$ 
      \\
      \rule[-1ex]{0pt}{3.5ex}  
      $\mathrm{cd}\ /\ \mathrm{nm}$ 
      & $[22.0 , 28.0]$ 
      & $25.48$ 
      & $0.59$ 
      & $ 0.1981$ 
      & $0.0188$ 
      & $5.49$ 
      \\
      \rule[-1ex]{0pt}{3.5ex}  
      $\mathrm{swa}\ /\ ^\circ$ 
      & $[84.0 , 90.0]$ 
      & $86.87$ 
      & $2.65$ 
      & $0.8847$ 
      & $-0.0172$ 
      & $2.14$ 
      \\
      \rule[-1ex]{0pt}{3.5ex}  
      $t\ /\ \mathrm{nm}$ 
      & $[ 4.0 ,  6.0]$ 
      & $ 4.96$ 
      & $0.35$ 
      & $0.3511$ 
      & $0.0074$ 
      & $2.93$ 
      \\
      \rule[-1ex]{0pt}{3.5ex}  
      $r_\mathrm{top}\ /\ \mathrm{nm}$ 
      & $[ 8.0, 13.0]$ 
      & $10.65$ 
      & $2.30$ 
      & $0.9202$ 
      & $-0.0466$ 
      & $2.14$ 
      \\
      \rule[-1ex]{0pt}{3.5ex}  
      $r_\mathrm{bot}\ /\ \mathrm{nm}$ 
      & $[ 3.0 , 7.0]$  
      & $ 4.89$  
      & $1.80$  
      & $0.8991$ 
      & $-0.0482$ 
      & $2.11$ 
      \\
      \hline
      \rule[-1ex]{0pt}{3.5ex}  
      $b$
      & $[ 0.0 ,  0.1]$
      & $0.01$
      & $0.0042$
      & $0.1698$ 
      & $1.0199$ 
      & $17.51$ 
      \\
      \hline
    \end{tabular}
  \end{center}
\end{table} 

The vector $\hat\xi$ consists of geometry parameters $\xi$ and the noise
parameter, i.e. $\hat\xi = (\xi,b)$. Assuming normal distributed zero-mean
measurement errors, we choose the likelihood function \cite{HGB15}

\begin{align}
  \label{eqn:likelihood}
  \mathcal{L}(\hat\xi;y)
  = \prod_{j=1}^d \frac{1}{\sqrt{2\pi}\sigma_j(b)}
  \exp\left( -\frac{(f^{(j)}(\xi)-y_j)^2}{2\sigma_j^2(b)} \right),
\end{align}

where $f^{(j)}$ is the $j$-th component of (the vector valued function) $f$.
Note that the form of the measurement error has to be chosen
appropriately not to introduce a bias. A more general approach in our
case would be not to impose a zero mean but introduce another random hyper
parameter. However, the residuum in Fig.~\ref{fig:reconstruction} suggests that
the noise is distributed around zero. In the Bayesian framework, the
distributions of parameters are in general determined by Markov-Chain Monte
Carlo (MCMC) sampling where for every sampling step, the forward model has to
be evaluated.  Normally, this means that \revB{equation
\eqref{eqn:Maxwell}} has to be solved which makes MCMC sampling impractical due
to the large number of required sampling steps. Since the surrogate only
requires evaluations of polynomials, the Bayesian approach becomes practical
for scatterometry measurement evaluations \cite{HGB18}.  

\begin{figure} [ht]
  \begin{center}
    \begin{tabular}{c} 
      \includegraphics[width=.9\linewidth]{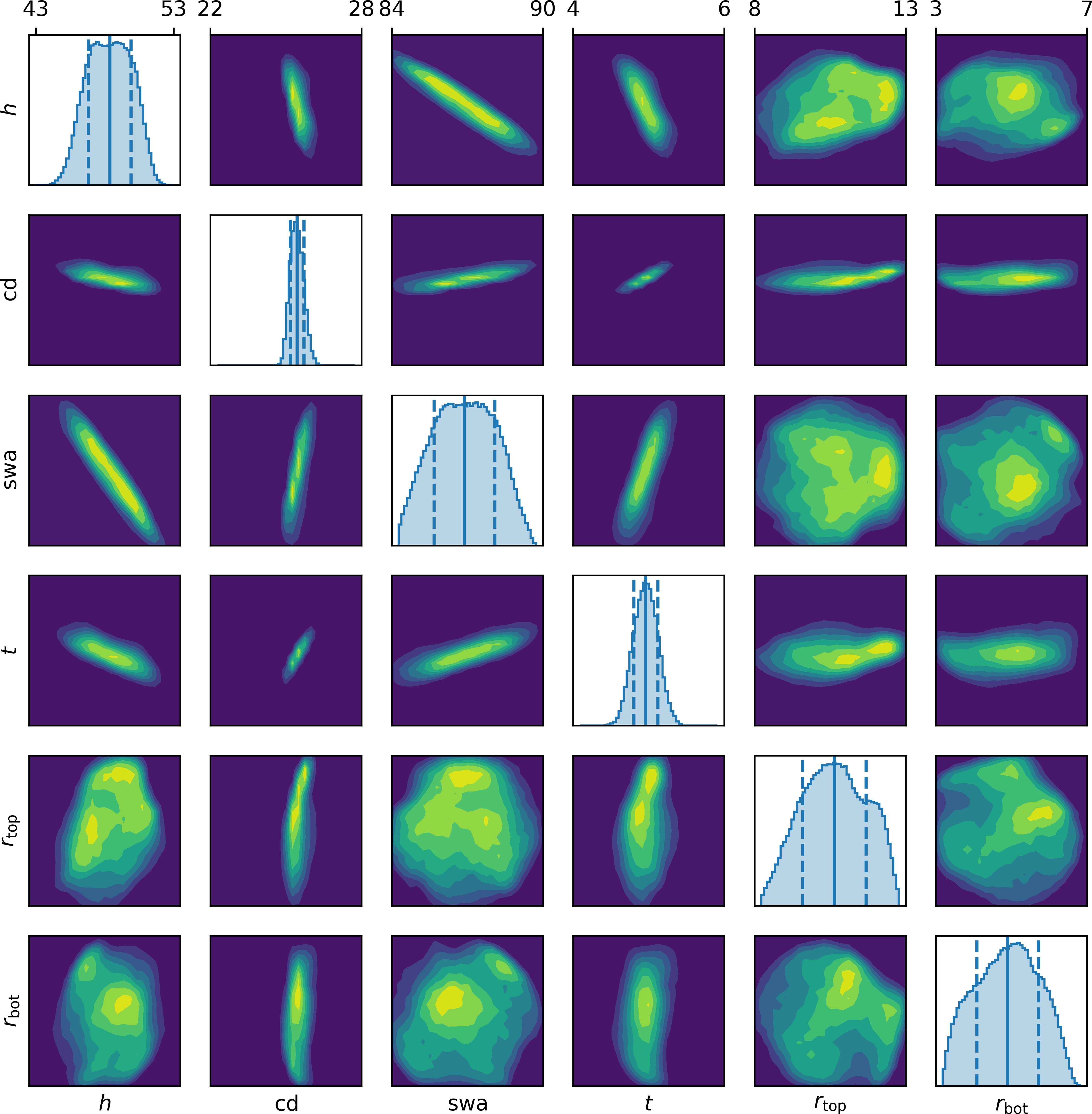}
   \end{tabular}
 \end{center}
 \caption{\label{fig:marginal_densities} 
 Marginal 1D and 2D densities for the posterior of the stochastic parameters.
 For the 1D densities, the mean (solid line) and the standard deviation 
 (dashed line) are depicted as well.
 }
\end{figure} 

For Bayesian inversion, we have to choose a prior distribution for the
parameters, calculate the likelihood function and determine the
corresponding posterior distribution. The posterior distribution contains the
desired parameter values and their associated uncertainties. When two or more
measurement results from different measurement sets $y^{(1)},\dots,y^{(K)}$ are
combined, the posterior distribution of the first measurement can be used as
the prior distribution for the evaluation of the second measurement, i.e.

\begin{align}
  \label{eqn:combined_measurements}
  \pi(\hat\xi;y^{(K)},y^{(K-1)},\dots,y^{(1)}) 
  = \frac{\pi_0(\hat\xi) \prod_{k=1}^K\mathcal{L}(\hat\xi;y^{(k)}) }
  {\int\pi_0(\hat\xi) \prod_{k=1}^K\mathcal{L}(\hat\xi;y^{(k)})
  \,\mathrm{d}\hat\xi}.
\end{align}

Note that the model function $f$ in the likelihood function is in general
different for different measurement setups.

\section{RESULTS}
\label{sec:results}

First we want to emphasize the efficiency of our approach. For the
scattering problem at hand, it is sufficient to use a chaos expansion with
$217$ terms to achieve a relative empirical $L^2$-error of less than $1\%$.
Therefore, in the sense of section~\ref{sec:sampling}, we generate
approximately $10^4$ samples for the FEM forward model to evaluate. In
comparison, the computation of the function mean, variance and Sobol indices or
the generation of posterior samples, if done empirically, require more than
$10^6$ function evaluations each due to the slow convergence rate of Monte
Carlo integration.

We apply Bayesian inversion to the scatterometry measurements to estimate
geometry parameters of the line grating. More details of the measurement setup
are described in previous works \cite{WBBR11,HWS+17}.  A global sensitivity
analysis for the geometry parameters \cite{FHS+19} indicates that the
reconstruction of all parameters is possible, i.e.  the forward model is
sensitive to all of them. In particular, the oxide layer thickness and
critical dimension should be possible to determine precisely due to their high
sensitivity. 

For Bayesian inversion it is necessary to chose prior distributions. In our
investigations we have chosen uniform priors on the domains given in
Table~\ref{tab:parameter}.  To obtain the posterior distribution, we sampled
with an MCMC random walk Metropolis-Hastings algorithm using the
surrogate.  We have chosen a sampling size of $10^6$ samples and a burn in
phase of $10^4$ samples. For diagnostics, we applied the Gelman-Rubin criterion
\cite{GR92}, to assure that the chains have fully explored the posterior.

\begin{figure} [ht]
  \begin{center}
    \begin{tabular}{c} 
      \includegraphics[width=.9\linewidth]{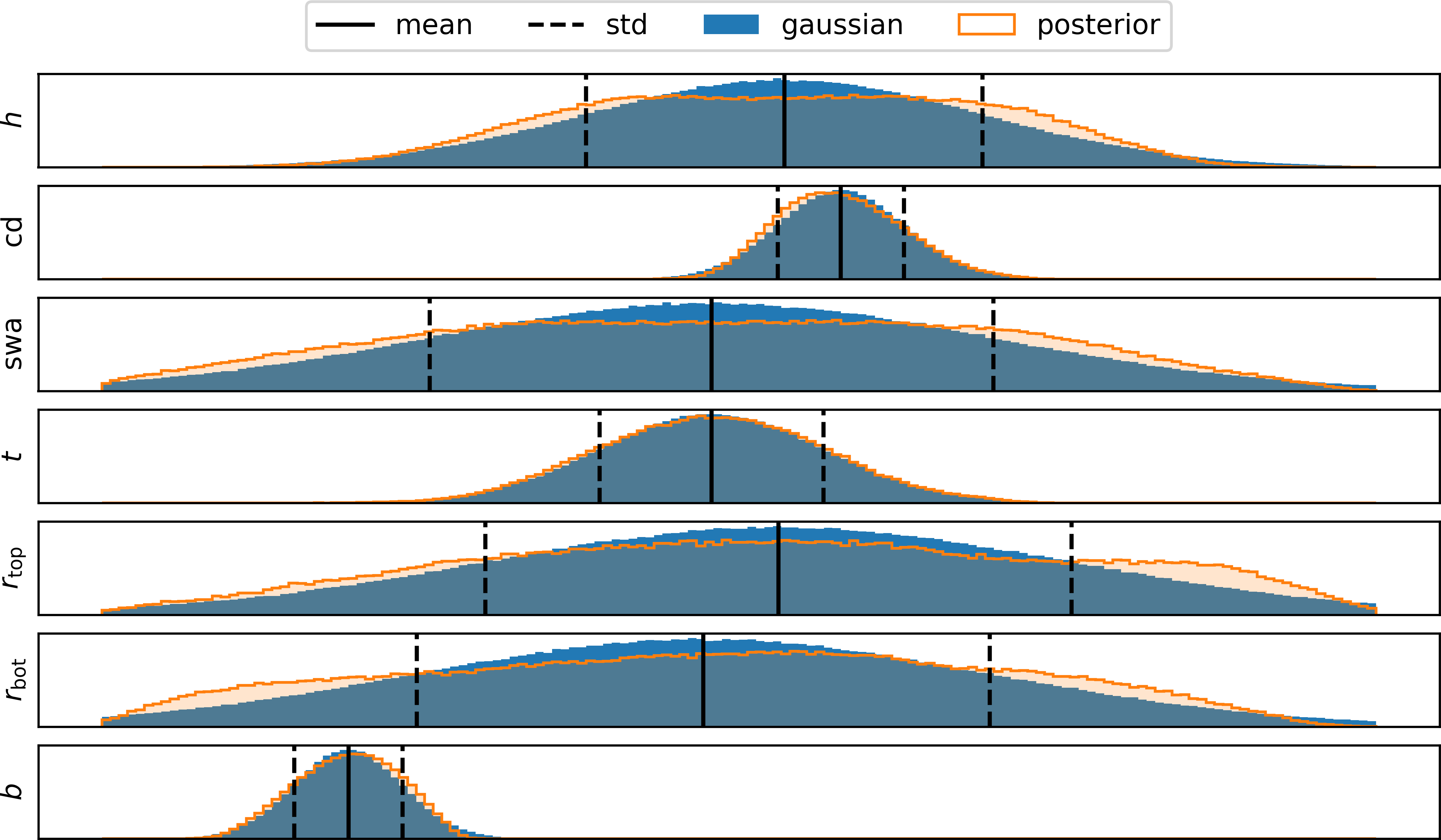}
   \end{tabular}
 \end{center}
 \caption{\label{fig:gauss_posterior} 
 Deviation of posterior marginals from Gaussian distribution with the same
 mean and variance. The distributions are depicted in their respective
 reconstruction domains (see Table~\ref{tab:parameter}).
 }
\end{figure}

In Fig.~\ref{fig:marginal_densities} the posterior (marginal) densities for all
$6$ stochastic parameters are shown.
All posterior densities are characterized by sharp peaks with mean
and standard deviation similar to the previous publication
\cite{HWS+17}. 
The mean and double standard deviation for each parameter including the
hyperparameter (error parameter) $b$ are shown in Table~\ref{tab:parameter}.
Since the domain sizes of the parameters vary due to their geometrical meaning,
we introduce the relative double standard derivation ($\relstd$).  The
$\relstd$ of a stochastic variable $\eta$ is the double standard deviation
divided by half the width of the parameter domain:

\begin{align}
  \label{eqn:relstd}
  \relstd = \frac{4\sigma}{\beta-\alpha}
  \qquad\text{where }\eta \in [\alpha,\beta].
\end{align}

The $\relstd$ shows how the posterior distribution is spread within the
domain.  For example, if the domain for the critical dimension is $[22,28]
\,\mathrm{nm}$, and the $2\sigma$ is $0.6\,\mathrm{nm}$ then the $\relstd$
is $0.198$, i.e. the posterior distribution is concentrated in about $20\%$ of
the originally chosen domain.  This way we can deduce how wide the parameter
distributions are spread across the reconstruction domains.  The $\relstd$
in Table~\ref{tab:parameter} shows that the smallest reconstruction
uncertainties are obtained for the critical dimension with $\relstd$ about
$20\%$, followed by the oxide layer thickness with $35\%$ $\relstd$. The
height has a $\relstd$ of about $62\%$. The posterior densities of the
sidewall angle and the corner rounding are slightly wider distributed at about
$90\%$ $\relstd$.  This goes in line with the global sensitivity analysis
\cite{FHS+19}.  The results for the error parameter $b$ depicted in
Table~\ref{tab:parameter} show that the relative measurement uncertainty is
approximately $1\%$. 

One major advantage of the Bayesian inference is information about the complete
posterior distribution instead of just  parameter values obtained from the
global minimizer. Looking at the marginals in
Fig.~\ref{fig:marginal_densities}, it is easy to verify that the posterior is
not Gaussian. The densities of the rounding radii are not symmetric, the
marginal distribution of the sidewall angle exhibits a plateau around
the mean and the height even suggests multi-modalities. Another validation of
these observations can be found in the skewness (third moment) and kurtosis
(forth moment) of the posterior. These differ (except for the oxide layer
thickness) quite significantly from the skewness and kurtosis of a Gaussian,
see Table~\ref{tab:parameter}. The deviation of the marginals from a Gaussian
with the same mean and standard deviation is shown in
Fig.~\ref{fig:gauss_posterior} for all parameters .

In our case the marginal distributions of the posterior are similar enough to a
Gaussian distribution that the $2\sigma$ confidence interval contains roughly
$95\%$ of the mass, as displayed in Table~\ref{tab:confidence}. However, in
general it is more reasonable to directly compute intervals of mass
concentration (confidence intervals) rather than relying on the standard
deviation to characterize the uncertainties of a distribution, because this can
be misleading for non-Gaussian distribution shapes occurring for example in
\cite{HGB18,FPPSS19}.

\begin{table}[ht]
  \caption{ \label{tab:confidence}
  Double standard deviation and $95\%$ mass confidence intervals of all
  geometry parameters and the error hyperparameter.
  }
  \begin{center}       
    \begin{tabular}{|p{2.5cm}p{2.5cm}p{3.8cm}|}
      \hline
      \rule[-1ex]{0pt}{3.5ex}
      parameter & $2\sigma$ interval & $95\%$ confidence interval \\
      \hline
      \rule[-1ex]{0pt}{3.5ex}  
      $h\ /\ \mathrm{nm}$ 
      & $(45.24,51.47)$ 
      & $(45.46,51.13)$ 
      \\
      \rule[-1ex]{0pt}{3.5ex}  
      $\mathrm{cd}\ /\ \mathrm{nm}$ 
      & $(24.88,26.07)$ 
      & $(24.95,26.09)$ 
      \\
      \rule[-1ex]{0pt}{3.5ex}  
      $\mathrm{swa}\ /\ ^\circ$ 
      & $(84.22,89.52)$ 
      & $(84.43,89.27)$ 
      \\
      \rule[-1ex]{0pt}{3.5ex}  
      $t\ /\ \mathrm{nm}$ 
      & $(4.61,5.31)$ 
      & $(4.61,5.30)$ 
      \\
      \rule[-1ex]{0pt}{3.5ex}  
      $r_\mathrm{top}\ /\ \mathrm{nm}$ 
      & $(8.35,12.96)$ 
      & $(8.51,12.67)$ 
      \\
      \rule[-1ex]{0pt}{3.5ex}  
      $r_\mathrm{bot}\ /\ \mathrm{nm}$ 
      & $(3.09,6.69)$  
      & $(3.26,6.49)$  
      \\
      \hline
      \rule[-1ex]{0pt}{3.5ex}  
      $b$
      & $(0.0054,0.0139)$
      & $(0.0056,0.0133)$
      \\
      \hline
    \end{tabular}
  \end{center}
\end{table} 

\begin{figure} [ht]
  \begin{center}
    \begin{tabular}{c} 
      \includegraphics[width=.7\linewidth]{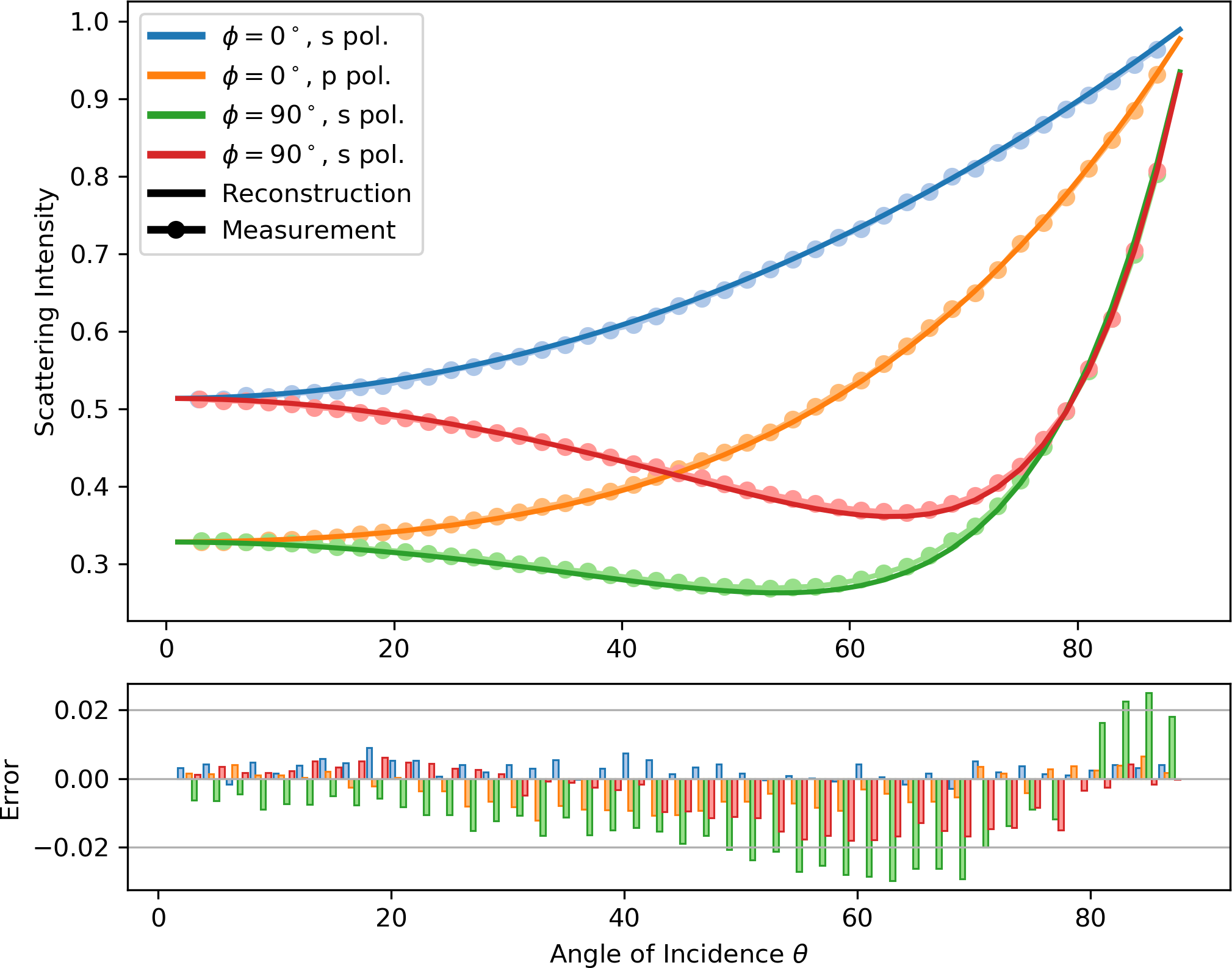}
    \end{tabular}
  \end{center}
  \caption{\label{fig:reconstruction} 
    Scattered intensities for the two polarizations and different azimuthal
    angles. Compared are the measurements of the scatterometry experiment
    and the simulation of the PC surrogate for the mean values of the 
    parameter reconstruction. The bottom graph shows the pointwise deviation.
    }
\end{figure}

Finally, Fig.~\ref{fig:reconstruction} displays a comparison between the
measurement data and the evaluation of our surrogate model using reconstructed
geometry parameters.  The pointwise relative deviation of the approximation
from the measurement data is $2\%$ and lower.  This is in accordance with
the reconstructed value of $b$, which describes the mismatch between the
surrogate of the forward model $f$ and the measurements. In a previous
work\cite{HWS+17} a Maximum Posterior Approach (MPA) incorporating the same
measurement data was used to determine the geometry parameters. The MPA
searches for the global maximum of the posterior. Uncertainties were determined
locally with an approximation of the covariance matrix around the maximum
posterior point. The difference here is that we calculated the whole posterior
distribution. This has the advantage that even for multiple peaked and
non-Gausian posterior distributions this scheme gives reliable uncertainty
estimations.  The results obtained in \cite{HWS+17} are consistent to our
findings since the posterior is relatively close to the assumed Gaussian shape.
There are only slight differences. For example, the marginal distribution for
the height $h$ is broad (non-Gaussian) yielding larger uncertainties.
Similarly, the mean values for $r_{\mathrm{top}}$ and $r_{\mathrm{bot}}$ are
slightly shifted due to the asymmetry of the marginal posterior (non-Gaussian).
The deviation between the forward model values and the measurement data of
$2\%$ is comparable with that found in \cite{HWS+17}.

\section{CONCLUSION}
\label{sec:summary}

In this paper we applied a polynomial chaos expansion as a surrogate for the
forward model in scatterometry. Since the surrogate
only requires the evaluation of polynomials instead of solving Maxwell's
equation, it was feasible to use a full Bayesian approach to determine the
posterior distribution for all geometry parameters. To generate samples from
the posterior distribution, we employed a MCMC Metropolis random walk sampling
method and checked the overall independence of the samples obtained by the
Gelman-Rubin criterion. The reconstruction results obtained by the surrogate
model compared to those obtained by a Maximum Posterior estimate with a
Gauss-Newton like method \cite{HWS+17} are consistent and are in line with the
predictions from a global sensitivity analysis. We conclude that a
Bayesian approach based on the polynomial chaos surrogate gives accurate and
reliable estimations for silicon line grating parameters and uncertainties.

\bibliography{report} 
\bibliographystyle{spiebib} 


\end{document}